\documentstyle[12pt,epsf,colordvi]{article}
\renewcommand{\baselinestretch}{1.3}

\def\listitem{\par \hangindent=50pt\hangafter=1
     $\ $\hbox to 20pt{\hfil $\bullet$ \hfil}}

\def\puncspace{\ifmmode\,\else{\ifcat.\C{\if.\C\else\if,\C\else\if?\C\else%
\if:\C\else\if;\C\else\if-\C\else\if)\C\else\if/\C\else\if]\C\else\if'\C%
\else\space\fi\fi\fi\fi\fi\fi\fi\fi\fi\fi}%
\else\if\empty\C\else\if\space\C\else\space\fi\fi\fi}\fi}
\def\SP{\let\\=\empty\futurelet\C\puncspace}

\def\h1{$h^{-1}$\SP}

\newcommand{\mincir}{\raise -2.truept\hbox{\rlap{\hbox{$\sim$}}\raise5.truept
\hbox{$<$}\ }}
\newcommand{\magcir}{\raise -2.truept\hbox{\rlap{\hbox{$\sim$}}\raise5.truept
\hbox{$>$}\ }}
\newcommand{\minmag}{\raise-2.truept\hbox{\rlap{\hbox{$<$}}\raise 6.truept\hbox
{$>$}\ }}

\def\lsim{~\rlap{$<$}{\lower 1.0ex\hbox{$\sim$}}}
\def\gsim{~\rlap{$>$}{\lower 1.0ex\hbox{$\sim$}}}
\def\void#1{{}}

\def\Red#1{\Color{0 1 1 0}{#1}}

\def\Cyan#1{\Color{1 0 0 0}{#1}}

\def\Magenta#1{\Color{0 1 0 0}{#1}}

\def\Green#1{\Color{1 0 1 0}{#1}}

\def\Blue#1{\Color{1 1 0 0}{#1}}

\def\Brown#1{\Color{.5 .5 1 0}{#1}}

\begin{document}

\begin{center}
{\Large \bf
\Blue {Optical to IRAS galaxy bias factor using the 
Local Group Dipole}}

\vspace{1cm}

\Red {\bf M. Plionis$^{1}$ \& S. Basilakos$^{1,2}$}

\vspace{1cm}

\Cyan {
$^{1}$ Astronomical Institute, National Observatory of Athens,  \\
I.Metaxa \& Bas.Pavlou, Lofos Koufou, 15236, Athens, Greece \\
$^{2}$ Section of Astronomy \& Astrophysics, Univ. of Athens,\\
Panepistimioupolis, 15784 Zografos, Athens, Greece
}
\end{center}

\vspace{0.8 cm}

\Brown {
\renewcommand{\baselinestretch}{1.5}
Comparing the gravitational acceleration induced on the Local Group by 
optical (SSRS2; da Costa et al 1994, 1998) and IRAS (1.2 Jy; Fisher et al 1995
and 0.6 Jy - QDOT; Rowan-Robinson et al 1990) galaxies 
we estimate, within the framework of linear theory, their relative bias 
factor. Using both IRAS samples we find $\Red {\bf b_{OI}\approx 1.14 - 1.2}$,
slightly lower than in Willmer, Da Costa \& Pellegrini (1998) who use 
a $\xi(r)$ approach.}

\vspace{0.6 cm}

\renewcommand{\baselinestretch}{1.3}
{

\noindent
Using linear perturbation theory one 
can relate the gravitational acceleration induced on the observer by the
surrounding mass distribution to her/his peculiar velocity:
$$
{\bf u(r)}=\frac{\Omega_{\circ}^{0.6}}{b} \frac{1}{4 \pi} \int 
\delta({\bf r})
\frac{\bf r}{|{\bf r}|^3} {\rm d}r = 
\frac{\Omega_{\circ}^{0.6}}{b} {\bf D}(r) 
$$
where 
$${\bf D} =\frac{1}{4 \pi \bar{n}} 
\sum_{i=1}^{N} \frac{1}{\phi_{i} r_{i}^{2}} \hat{{\bf r}}_i$$
with
$\phi(r)=\langle n \rangle^{-1} \int_{L_{{\rm min}}(r)}^{L_{{\rm max}}} 
\Phi(L)\,{\rm d}L$,
where $\Phi(L)$ is the luminosity function of the objects under study and
$L_{\rm {min}}(r)=4\pi r^{2} S_{\rm {lim}}$, with $S_{\rm {lim}}$
being the flux limit of the sample under study.

\noindent
Different classes of
extragalactic objects (QSOs, AGNs, galaxies, clusters of galaxies) 
trace differently the underlying matter distribution, usually assumed to be
linearly related via $\left(\delta \rho/\rho \right)_{\rm tracer} = 
b_{\rm tracer,m} \left(\delta \rho/\rho \right)_{\rm mass}$
(cf. Kaiser 1984). Although this complicates the use of dipole 
as an estimator of the cosmological density parameter, $\Omega_{\circ}$,
it does allow one in principle to study the relative bias displayed
by such objects (Kolokotronis {\em et al} 1996; Plionis 1995). 
Using different tracers, ie. SSRS2 optical and IRAS galaxies, to determine 
the Local Group acceleration (dipole) we can write:
${\bf u(r)}= \Omega_{\circ}^{0.6} {\bf D_{o}}(r)/b_{o} = 
\Omega_{\circ}^{0.6} {\bf D_{I}}(r) /b_{I}$
and therefore we can obtain an estimate of their relative bias factor from:
$$ b_{OI}=\frac{b_{o}}{b_{I}} = \frac{{\bf D}_{o}(r)}{{\bf D}_{I}(r)}
\left( \equiv \frac{{\bf g}_{o}(r)}{{\bf g}_{I}(r)} \right)$$
A statistically more reliable approach is to use the differential dipole, 
estimated in equal volume shells, to fit $b_{OI}$ according to:
$$ \chi^2 = \sum_{i=1}^{N_{bins}} \frac{({\bf D}_{o,i} - b_{OI} {\bf D}_{I,i} 
- C_{i})^{2}}{\sigma_{o,i}^2 + b_{OI}^{2} \sigma_{I,i}^2}$$
where $C$ is the zero-point offset of the relation and $\sigma$ are
the corresponding shot-noise errors estimated according to Strauss et al. 
(1992).

\begin{figure*}[h]
\mbox{\epsfxsize=13cm \epsffile{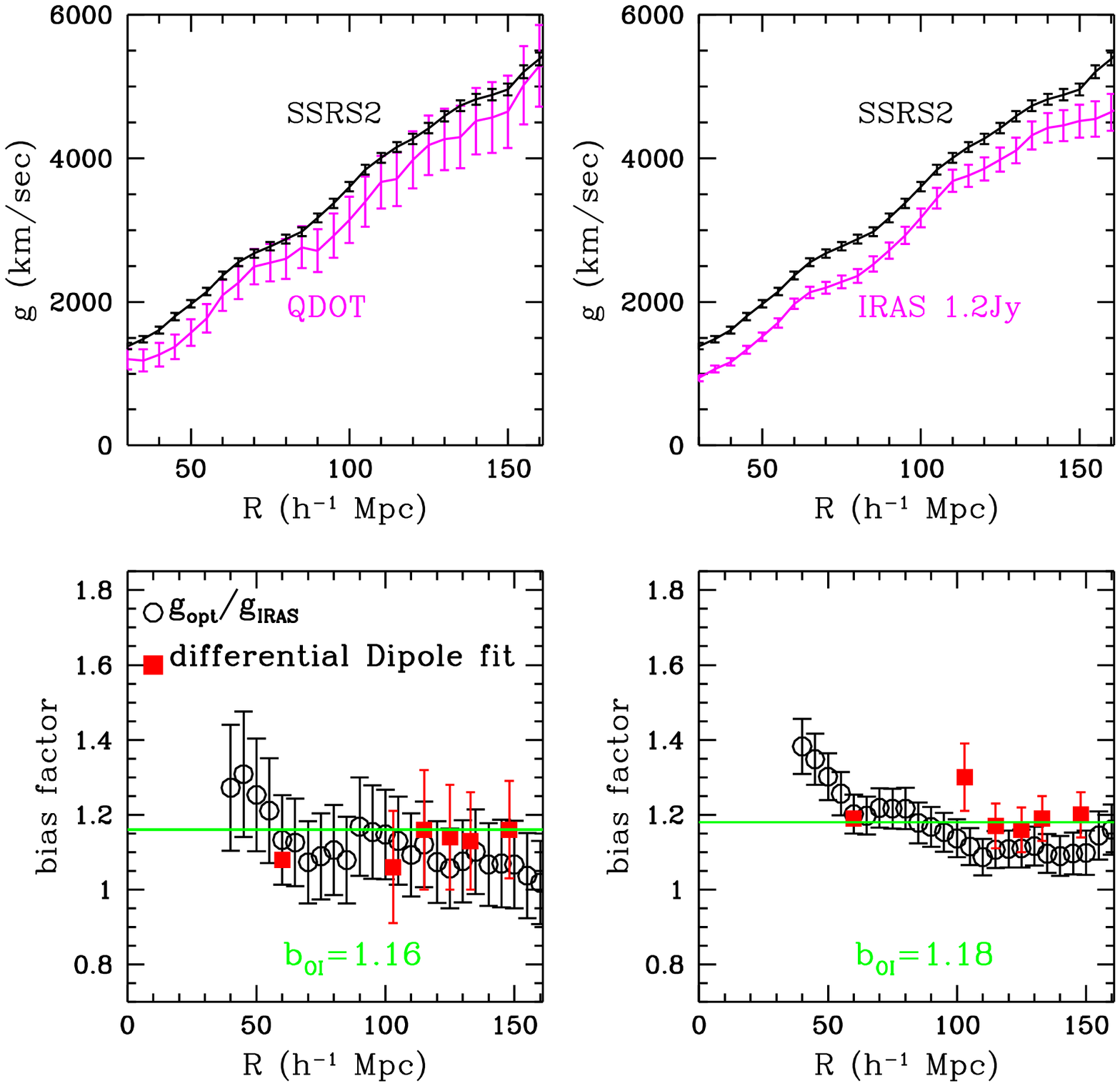}}
\end{figure*}

\noindent
In the figure and in table 1 we present the results of both methods used to 
estimate the optical (SSRS2) to infrared (IRAS 1.2 Jy and QDOT) bias factors.
}
\begin{table}[h]
\caption[]{Optical to Infrared galaxy bias factors from differential dipole
fit.}
\tabcolsep 18pt
\Blue {
\begin{tabular}{lcccc}
samples & $b_{OI}$ & $C$ & $\chi^2$ & ${\rm d.f.}$ \\ \hline 
\Brown {SSRS2-QDOT} & $\Red {\bf 1.16 \pm 0.13}$ & -80 $\pm 70$ & 8.7 & 6 \\
\Brown {SSRS2-IRAS 1.2 Jy} & $\Red {\bf 1.18 \pm 0.06}$ & -30 $\pm 30$ & 15.7 & 6 \\ \hline
\end{tabular}
}
\end{table}

\vspace{1cm}

{\small 
\section*{References}

\noindent 
da Costa, L.N. {\em et al.}, 1994, ApJ, 424, L1

\noindent 
da Costa, L.N. {\em et al.}, 1998, {\em in preparation}

\noindent
Fisher, K.B., Huchra, J.P., Strauss, M.A., Davis, M., Yahil, A. 
\& Schlegel, D., 1995, ApJS, 100, 69 

\noindent
Kaiser N., 1984, ApJ, 284, L9 

\noindent
Kolokotronis, V. {\em et al.}, 1996, MNRAS, 280, 186

\noindent
Plionis, M., 1995, in Balkowsky C. {\em et al}, eds, Proc. of the 
Moriond Astrophysics Meeting on Clustering in the Universe, Editions 
Frontieres.

\noindent
Rowan-Robinson M., {\em et al.}, 1990,  MNRAS, 247, 1 

\noindent
Strauss M., Yahil A., Davis M., Huchra J.P., Fisher K., 1992
ApJ, 397, 395

\noindent
Willmer, C.N.A., Da Costa, L.N. \& Pellegrini, P.S., 1998, {\em astro-ph/9803118}

}

\newpage

\begin{center}
{\large \bf
\noindent
\Blue {The X-ray Luminosity Function of Local Galaxies}}

\vspace{0.75cm}

\Red {\bf S. Basilakos$^{1,2}$, I. Georgantopoulos$^{1}$, M. Plionis$^{1}$, O. Giannakis$^{1}$}

\vspace{0.5cm}

\Cyan {
$^{1}$ Astronomical Institute, National Observatory of Athens,  \\
I.Metaxa \& Bas.Pavlou, Lofos Koufou, 15236, Athens, Greece \\
$^{2}$ Section of Astronomy \& Astrophysics, Univ. of Athens,\\
Panepistimioupolis, 15784 Zografos, Athens, Greece
}
\end{center}

\vspace{0.5 cm}

\renewcommand{\baselinestretch}{1.4}
\Brown {
We construct  the X-ray luminosity (LF) 
for different classes of galaxies (Seyferts, ellipticals, 
 star-forming galaxies and Liners), 
  by convolving the optical LF of the Ho {\em et al.} 
spectroscopic sample of nearby galaxies with the 
corresponding $\Blue {L_x/L_o}$ relations from the Fabbiano et al. 
X-ray atlas of galaxies. From the derived LF we can easily assess the
 contribution of galaxies to the X-ray background. 
The \Red{Seyferts and Liners make the largest contribution ($\sim 40\%$) 
assuming no evolution} 
 while the contribution of star-forming galaxies is much smaller ($\sim 5\%$).}

\renewcommand{\baselinestretch}{1.3}
{
\begin{center}
{ {\Red {The Sample \& Method}} }
\end{center}
We have used the spectroscopic sample of nearby galaxies (B$<$12.5) 
of Ho {\em et al.} (1995). The great advantage of this sample is that there is 
excellent spectroscopic infomation (high signal-to-noise, medium 
resolution) available and thus bona-fide spectroscopic identifications 
exist for all ($\sim$ 500) galaxies in the sample. 
Hence, we can construct the X-ray LF separately for 
Seyferts, star-forming galaxies (HII), ellipticals and Liners 
instead of  simply dividing them to early-type and late-type 
according to their morphology. 
The majority of galaxies in the Ho {\em et al.} sample 
are \Red {HII (50\%), Liners are (30\%), Seyferts (13\%) 
while 15\% of galaxies present no emission lines} 
and thus can be classified as 'early-type' galaxies or
'ellipticals'on the basis of their spectra rather than their morphology
(Ho {\em et al.} 1997). We first derive the optical 
LF for different  classes of objects (Seyferts, HII, Liners
and no-emission-line or early-type galaxies) 
using the classical $\Red{1/V_{max}}$ method. 
Next, we derive the $L_x/L_B$ relation for the different subclasses using 
the {\small EINSTEIN} 
 X-ray fluxes (0.5-4.0 keV) from the Fabbiano et al. (1992) 
X-ray atlas of galaxies.
There are 164 entries (95 detections and 69 3$\sigma$ upper limits)
 of Ho et al. galaxies in the Fabbiano atlas.   
Finally, in order to derive the X-ray LF we convolve the optical 
LF with the $L_x/L_B$ relation:

$$
\Magenta {\Phi(L_x)= \int \Phi(L_B) \phi(L_x| L_B) dL_B}
$$

\noindent 
(eg Avni \& Tananbaum 1986), 
where $\phi(L_x| L_B)$ is the conditional probability function 
and can be expressed as a Gaussian around the mean $L_x$ value 
for a given $L_B$ (we use $H_o=100$ throughout). 

\begin{figure*}[h]
\mbox{\epsfxsize=9cm \epsffile{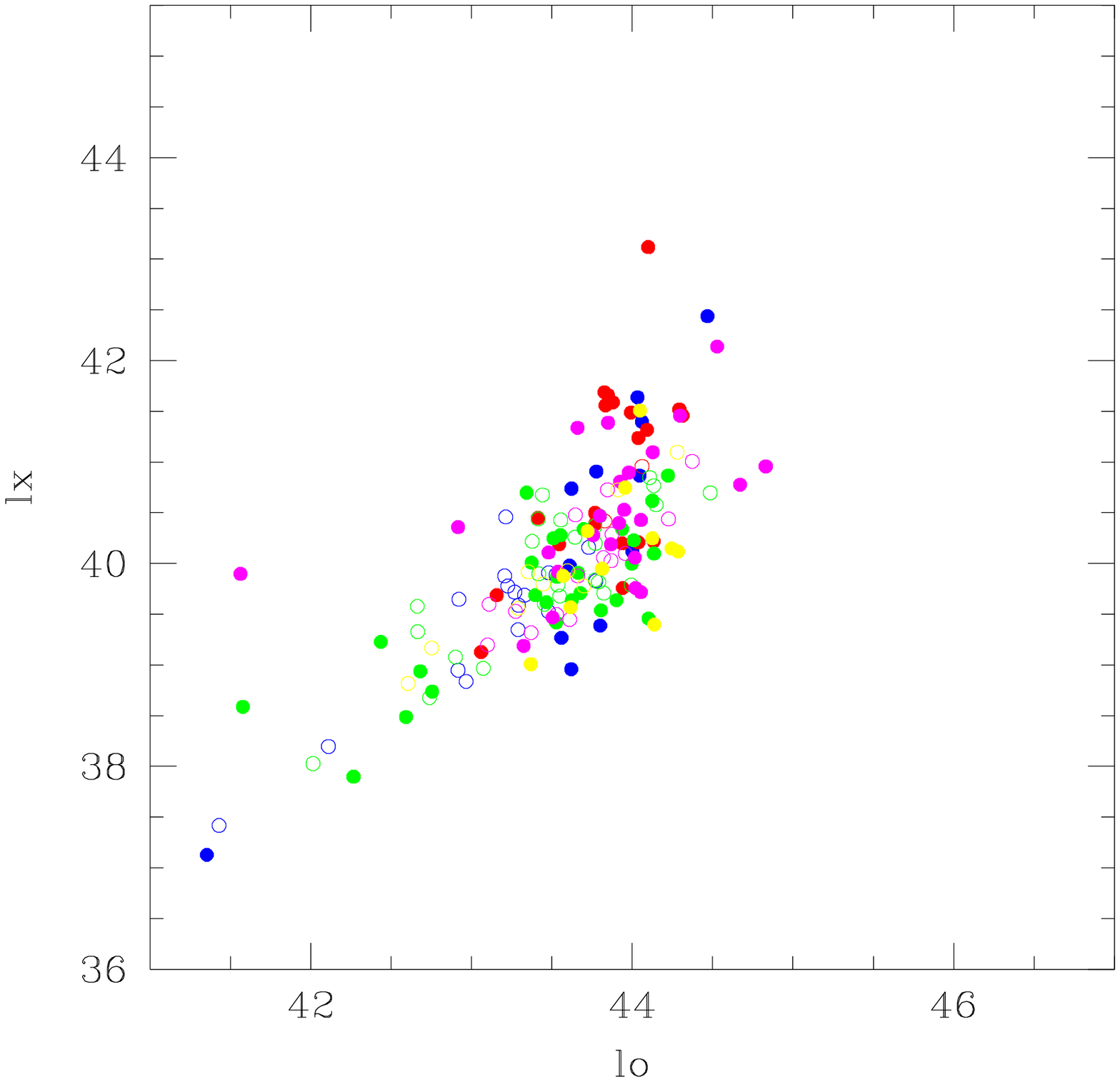}}
\end{figure*}

\noindent 
Although, our method provides only an indirect way of 
deriving the X-ray LF, it is  currently the only feasible method
at least for some classes of galaxies. Unfortunately, the 
galaxies in X-rays are faint (apart from  Seyferts) 
 and thus we cannot yet obtain large X-ray selected  
galaxy samples neither in deep X-ray surveys nor in 
the ROSAT all-sky survey. 
For example we note that there are only 5 HII galaxies 
from the Ho et al. sample detected by the RASS (Zezas {\em et al.} 1998). 

\begin{figure*}[h]
\mbox{\epsfxsize=10cm \epsfysize=8.5cm \epsffile{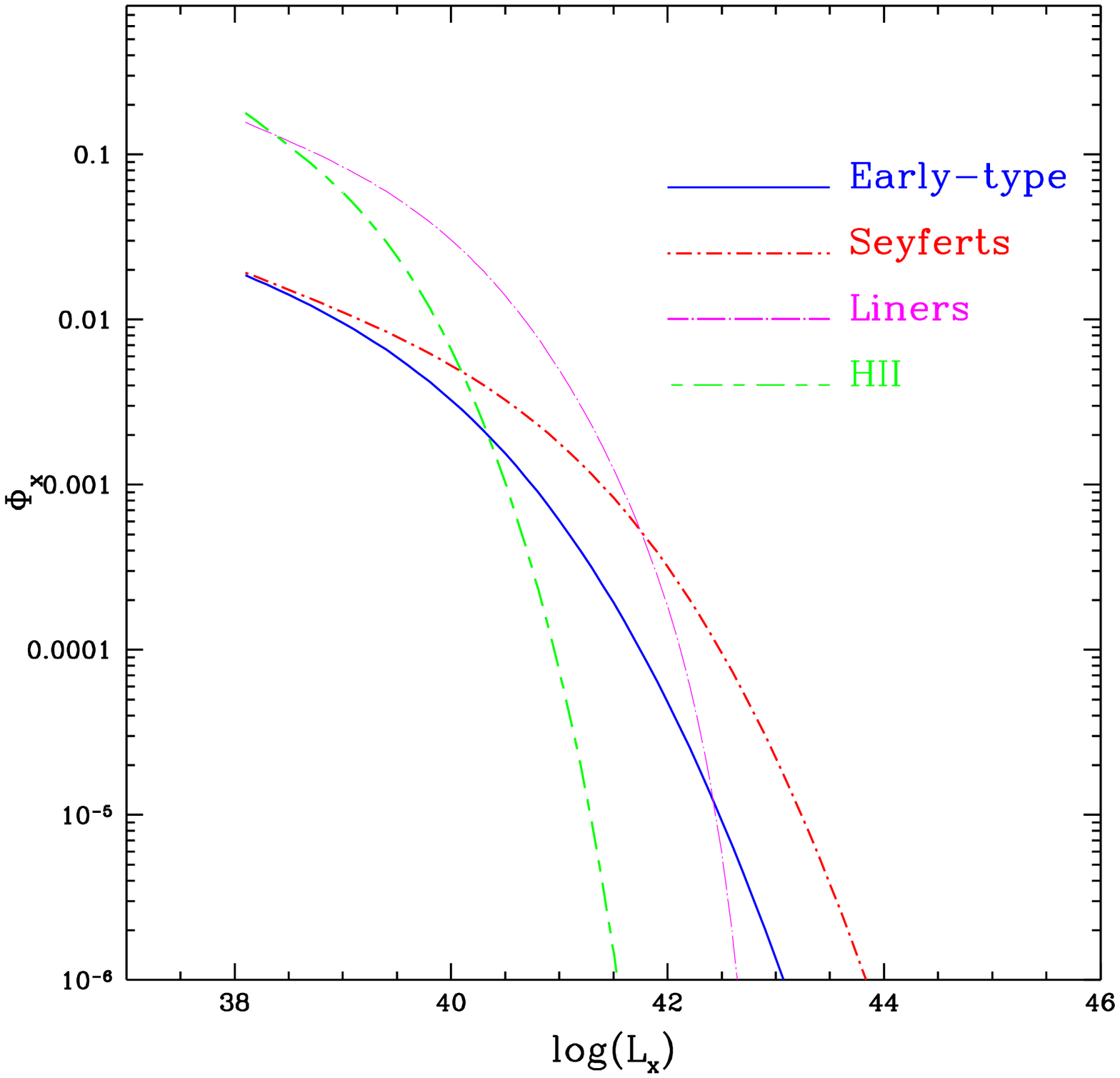}}
\end{figure*}

{
\begin{center}
{\Red { The Results}}
\end{center}
In the figure we plot the X-ray luminosities vs. the Blue luminosities
($\log_{10}(L_x)$ vs $\log_{10}(L_B)$), 
for different subclasses (open symbols represent upper limits while 
filled symbols represent detections; the color coding is as in the table).

\begin{table}
\caption{The best-fit $\Phi_x$ results for a Schechter optical luminosity 
function}
\tabcolsep 20pt
\begin{tabular}{ccccc}
Type & $\Phi_\star$ &  $\alpha$   &  $M_\star$ & $\chi^2/dof$  \\ \hline 
\Red{Seyferts &0.001&$-1.39^{+0.350}_{-0.460}$&$-20.09^{+0.290}_{-0.130}$& 
0.250}\\
\Magenta{LINER &0.005& $-1.19^{+0.230}_{-0.320}$ &$-19.77^{+0.200}_{-0.290}$ 
& 0.646} \\
\Green{HII &0.005 &$ -1.39^{+0.154}_{-0.088}$  & $-19.75^{+0.240}_{-0.376}$ 
&1.000} \\
\Blue {early-type &0.002&$-1.39^{+0.380}_{-0.560}$ &$-19.73^{+0.290}_{-0.370}$
&1.622}\\
All &0.016&$-1.29^{+0.040}_{-0.130}$& $-19.70^{+0.047}_{-0.500}$ &1.970  \\
\hline
\end{tabular}
\end{table}
 
\noindent
The X-ray LF is plotted on Fig. 2. It is clear that 
although the HII galaxies are the most numerous, due to their low $L_*$
 they contribute much less X-ray luminosity per $\rm h^{-3} \; Mpc^3$ 
compared to Seyfert galaxies.
Indeed, the HII emissivity is  $\Green{J\sim 1\times 10^{38}$ 
 $\rm erg~s^{-1} \; h^{3} \; Mpc^{-3}}$ 
 while the Seyfert is $\Red{J\sim 8\times 10^{38}$   
 $\rm erg~s^{-1} \; h^{3} \; Mpc^{-3}}$ . 
 The total galaxy emissivity is comparable to the emissivity of local galaxies
 in the 2-10 keV band derived 
by Lahav et al. (1993) from the cross-correlation of optical  
 galaxy catalogues with the fluctuations of the hard  
X-ray background. 
}

\newpage

\begin{center}
{\large \bf
\noindent
\Blue {The ROSAT X-ray Background Dipole}}

\vspace{0.75cm}

\Red {\bf M. Plionis \& I. Georgantopoulos}

\vspace{0.5cm}

\Cyan {
Astronomical Institute, National Observatory of Athens,  \\
I.Metaxa \& Bas.Pavlou, Lofos Koufou, 15236, Athens, Greece
}
\end{center}

\vspace{0.8 cm}

\Brown {
\renewcommand{\baselinestretch}{1.4}
We estimate the dipole of the diffuse 1.5 keV X-ray background from 
the ROSAT all-sky survey map of Snowden et al (1995).
We first subtract the diffuse  Galactic emission  by 
fitting to the data a finite disk model, following Iwan et al (1982). 
 We further exclude regions of low galactic latitudes,
of local X-ray emission (eg the North Polar Spur) and model
 them  using two different methods.
We find that the ROSAT X-ray background dipole points towards 
$\Red {\bf (l,b) \approx (288^{\circ}, 25^{\circ}) \pm 19^{\circ}}$  
within $\Red {\bf \sim 30^{\circ}}$ of the CMB; its direction is also in 
good agreement with the HEAO-1 X-ray dipole at harder energies.  
The normalised amplitude of the ROSAT XRB dipole is  $\Red {\bf \sim 1.6\%}$.}

\vspace{1.0 cm}

\renewcommand{\baselinestretch}{1.3}
{
\begin{center}
{\Red {Subtracting the Diffuse Galactic Emission}}
\end{center}
A major problem in extracting the X-ray background dipole at soft
X-rays is the large contribution of our Galaxy at such energies
(cf. Kneissl et al 1998).
We attempt to model the diffuse Galactic component using 
a finite radius disk with an exponential scale height, (Iwan et al. 1982)   
which provides a good description of the Galactic component at harder 
energies (2-60 keV). 
The fit to the model is performed by excluding the regions of the most apparent
extended Galactic emission features: the North Polar Spur and the 
$|b|< 20^{\circ}$ strip as well as most apparent "local" extragalactic 
features; a 
$4^{\circ}$ radius region around the Virgo cluster ($l,b \approx 287^{\circ},
75^{\circ}$) and a 10$^{\circ}$ radius region around the Magellanic clouds  
($l,b \approx 278^{\circ}, -32^{\circ}$).

\Magenta{We obtain a fraction of the total X-ray emission which 
is due to the Galaxy consistent with $\Red {\sim 25\%}$ while
the disk scale height and disk radius where found to be \Red{16} \& \Red{27 
kpc} respectively.}

\newpage

{
\begin{center}
{\Red {Dipole Estimation}}
\end{center}
After excluding from the ROSAT counts our best fit Galactic model and after
masking, using either of two methods; \Magenta {\em homogeneous filling 
procedure} or a \Magenta{\em spherical harmonic} extrapolation procedure, 
the following regions:
\Red{(a)} \Blue{ The Galactic plane, with $|b|\le 20^{\circ}$ or $30^{\circ}$},
\Red{(b)} \Blue{ the area dominated by the Galactic bulge and the North 
Polar Spur (ie., $-40^{\circ} < b < 75^{\circ}$ and $300^{\circ} < l 
< 30^{\circ}$)} ,
\Red{(c)} \Blue {the Large Maggelanic Clouds (ie., an area of 
10$^{\circ}$ radius centred on $l,b \approx 278^{\circ}, -32^{\circ}$)},
we measure the dipole by weighing the unit 
directional vector pointing to each 40$^2$ arcmin$^2$ ROSAT cell with
the X-ray intensity ${\cal C}_i$ of that cell. We normalize the dipole by the 
monopole term (the mean X-ray intensity over the sky):
$
{\cal D} \equiv |{\bf D}|/M = \sum {\cal C}_i \hat{\bf r}_i/
\sum {\cal C}_i $

\begin{center}
{ \Red {Dipole Results}}
\end{center}
When using the raw ROSAT data, the dipole points
towards the Galactic centre (in agreement with the analysis of
Kneissl et al 1997). However, when we exclude both the Galaxy and the North 
Polar Spur, the measured dipole is in much better directional agreement with
the CMB dipole. For the {\em homogeneous filling} method  we find
$\delta\theta_{cmb}\sim 20^{\circ}$ while for the {\em 
spherical harmonic} method the misalignment angle is larger, 
$\delta\theta_{cmb}\sim 51^{\circ}$.

\noindent
However, it should be expected that many Galactic sources, probably 
dominating the \Magenta {higher ROSAT counts}, are still present in 
the data and could affect the behaviour of the estimated extragalactic XRB 
dipole. 
Excluding the highest $\sim 3\%$ of the counts
($\Blue{ {\cal C}_{up} \magcir 140}$) we find that
{\Magenta{\em both methods used to model the masked areas \underline{agree} 
and the XRB-CMB dipole misalignment angle is reduced significantly} }

\noindent
The interpretation that the high intensity cells are 
associated with Galactic sources is supported by the fact that when we 
include these few cells the resulting dipole direction moves 
towards the Galactic centre. 

\noindent
Therefore, taking into account the variations of our results due to \Blue{(a)}
the uncertainties of the Galactic model subtracted from the raw counts, 
\Blue{(b)} to the different methods used to mask the excluded sky regions 
and \Blue{(c)} to the 
different galactic latitude limits, we conclude that the {\em extragalactic} 
ROSAT dipole has:
$$\Red{ {\cal D}_{\rm XRB,ROSAT} \approx 0.016 \pm 0.008 \;\;\;\;\; (l,b) 
\approx (288^{\circ}, 25^{\circ}) \pm 19^{\circ} }$$
Our  results are consistent with the HEAO-1 (2-10 keV) dipole 
(Shafer \& Fabian
1983) which points in a  similar direction ($282^{\circ}, 30^{\circ}$)
but has a lower amplitude: ${\cal D}_{\rm XRB,HEAO-1} \sim 0.005$.

\begin{figure*}[h]
\mbox{\epsfxsize=18cm \epsfysize=10cm \epsffile{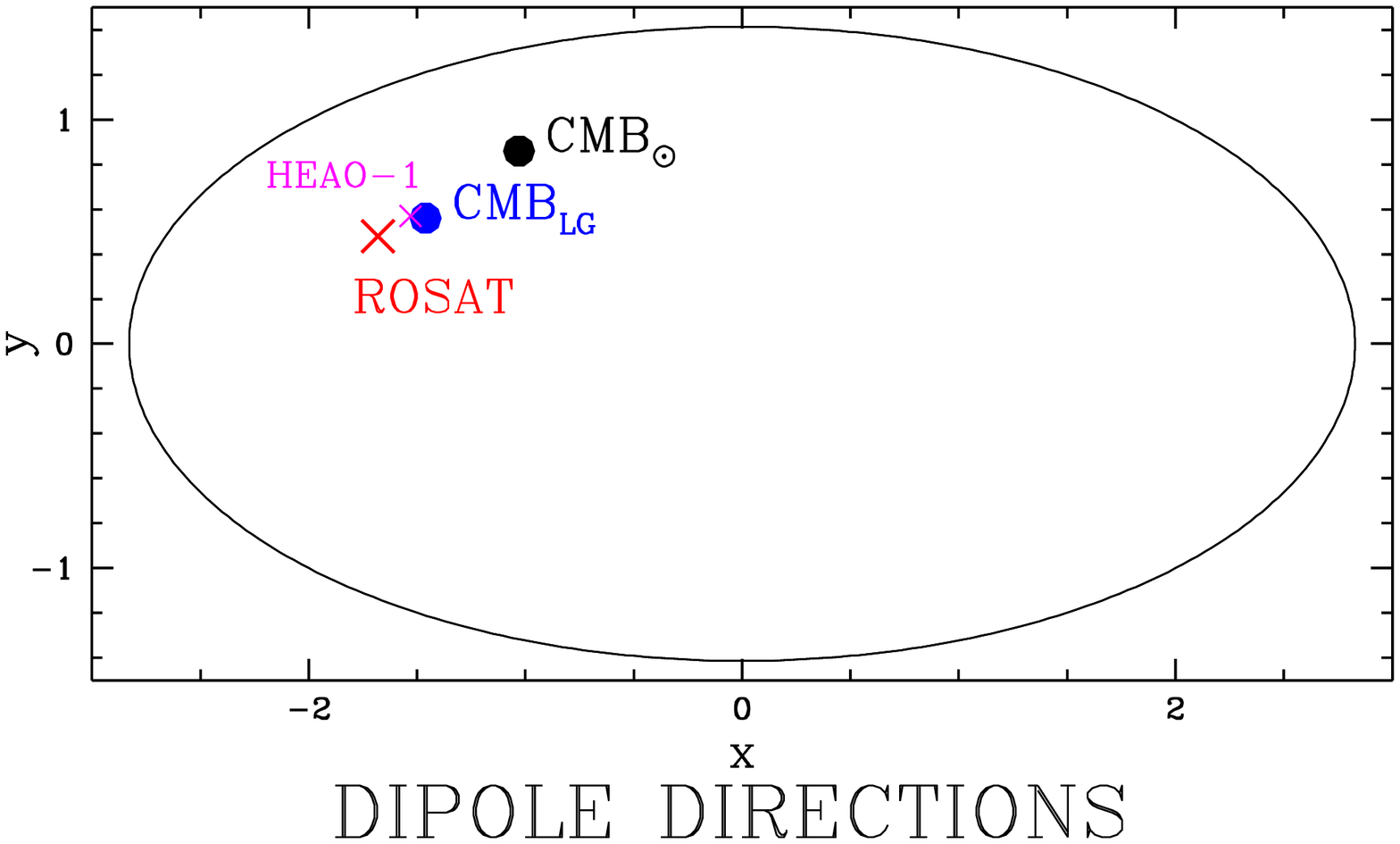}}
\end{figure*}

\newpage

\begin{center}

{\large \bf
\noindent
\Blue {The Angular Correlation Function of RASS Extragalactic Sources}}

\vspace{0.75cm}

\Red {\bf T. Akylas, M. Plionis \& I. Georgantopoulos}

\vspace{0.5cm}

\Cyan {
Astronomical Institute, National Observatory of Athens,  \\
I.Metaxa \& Bas.Pavlou, Lofos Koufou, 15236, Athens, Greece
}
\end{center}

\vspace{0.5 cm}

\renewcommand{\baselinestretch}{1.4}
\Brown {
We investigate the clustering properties of a new homogeneous sample of 2000 
extragalactic sources selected from the ROSAT XRT/PSPC All-Sky survey
(Voges {\em et al} 1996). Assuming that this sample is dominated by QSO's we find 
that the characteristic depth of the sample is $\sim 410 \; h^{-1}$ Mpc.
We find a \Blue{significant angular correlation function}, between 
$\sim 2^{\circ}$ and 15$^{\circ}$ which roughly corresponds to a 
spatial correlation length of $\Red{r_{\circ} \simeq 15.5 \pm 4.5 \; h^{-1}}$ 
Mpc, roughly consistent with that of optically selected QSO's at 
$\langle z \rangle \simeq 1.5$.}

\renewcommand{\baselinestretch}{1.3}
{
\begin{center}
{\Red {Method}}
\end{center}
The original RASS sample contains 18811 sources which include stars and 
extragalactic sources over the entire 0.1 - 2 keV energy range. 
To exclude the stellar objects we cross correlated this sample with star 
catalogs (SAO, GSC, RASSOB, RASSWD, XRBCAT and CVCAT). 
Furthermore, we have cross-correlated the remaining sources (for $b>0^{\circ}$)
with the \underline{Hamburg identifications} (Bade {\em et al} 1998) 
to find that most of them, 
after excluding those with extension flag \Blue{$>$ 40}, are identified as 
\Red{QSO's}.
In order to produce a homogeneous sample in exposure time we have omitted
sources with count rate $<$ 0.1 which results in a 
\Magenta{homogeneous sample over 92\% of the sky}.
Finally, we exclude regions heavily affected by Galactic absorption, ie.,
$\Blue{|b|<30^{\circ}}$ as well as $\Blue{\delta < -30^{\circ}}$ (see below).
\Magenta{Our final catalogue contains 2000 extragalactic sources}.

\noindent
We further need to estimate the various selection functions that could affect 
the clustering properties of our survey. For example the \Red{significant 
absorption 
due to the diffuse Galactic Neutral Hydrogen could artificially 
enhance the 2-point correlation function} and also introduce large-scale 
modulations. Therefore, we have cross correlated our RASS catalogue with
the recent Leiden/Dwingeloo Atlas of Galactic Neutral Hydrogen (Hartmann \& 
Burton 1997) which covers regions with $\delta\ge -30^{\circ}$, to derive 
the RASS surface density as a function of $N_{H}$, which is 
used to produce a random catalogue with similar absorption selection function.

{
\begin{center}
{\Red { Results}}
\end{center}
The correlation function estimator we use is:
$\Blue{ w(\theta) = f N_{dd}/N_{dr} -1}$,
where the normalization factor is $f = 2 \times N_{r}/N_{o}$.
Assuming a 2-p correlation function of the form 
$\Blue {w(\theta)=(\theta/\theta_{\circ})^{b}}$ we find 
$\Red {b=-0.9 \pm 0.15}$ and
$\Red{\theta_{\circ}=0.08^{\circ} \pm 0.05^{\circ} }$.

\noindent
A preliminary calculation using the average flux limit ($\sim 10^{-12}$ 
erg/sec cm$^{2}$) of the RASS over the area used 
and a QSO characteristic
luminosity of $L_{*} \simeq 2 \times 10^{43}$ (Boyle {\em et al} 
1993) gives a characteristic depth of $D_{*} \sim 410 \; 
h^{-1}$ Mpc. Using Limber's equation to relate the angular and spatial 2-point
correlation function with $\gamma \simeq 2$ we find a correlation length of 
$\Red {r_{\circ} \simeq 15.5 \pm 4.5 \; h^{-1} \; {\rm Mpc}}$,
 which is compatible only with comoving QSO 
clustering (cf. Shanks \& Boyle 1994 and La Franca {\em et al} 1998 which find 
$r_{\circ} \simeq 10 \pm 2.5 \; h^{-1}$ Mpc at 
$\langle z \rangle \simeq 1.5 - 1.8$)

\begin{figure*}[h]
\mbox{\epsfxsize=15cm \epsfysize=10cm \epsffile{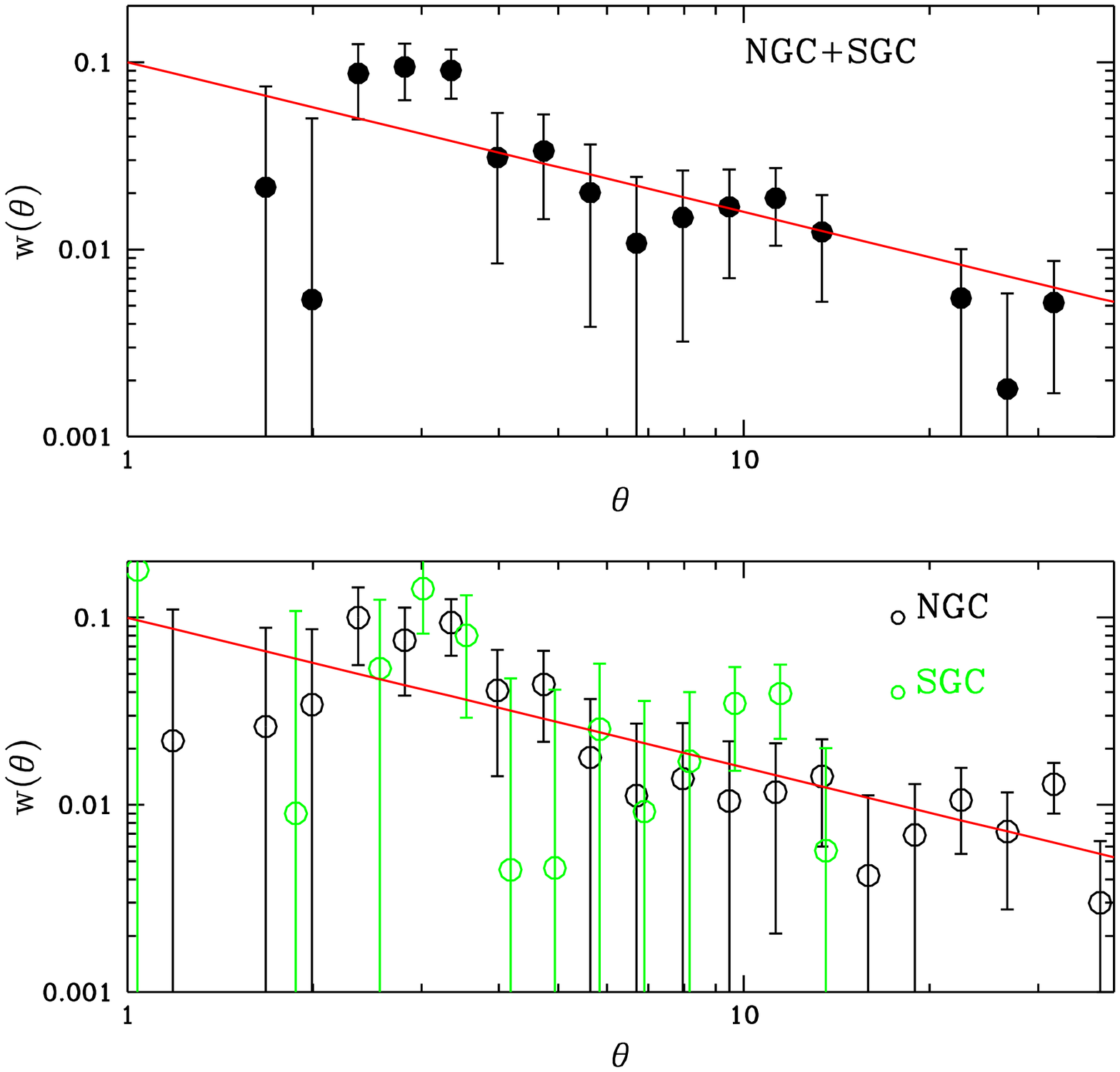}}
\end{figure*}

\renewcommand{\baselinestretch}{1.}
\begin{figure*}[h]
\mbox{\epsfxsize=16cm  \epsffile{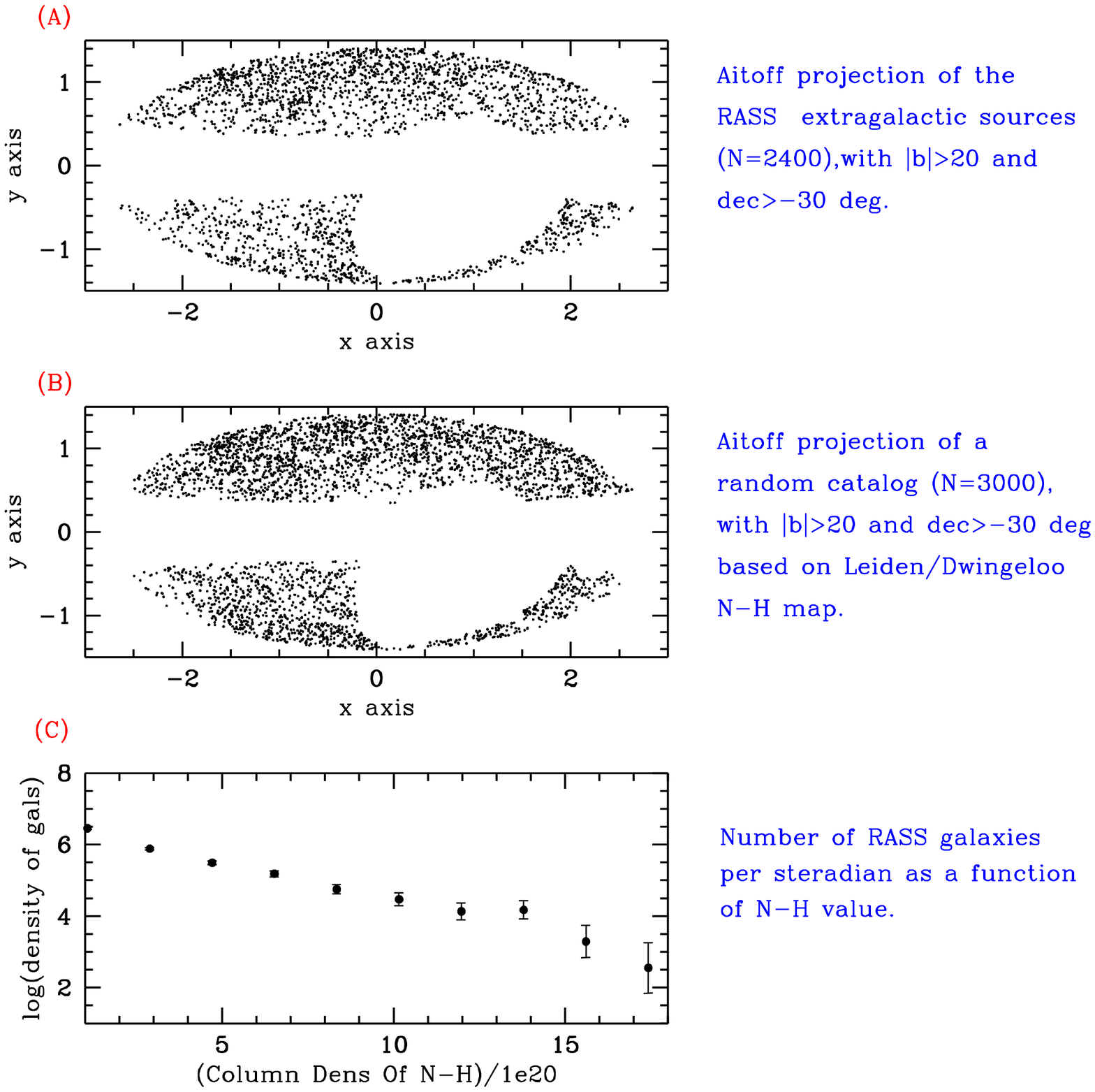}}
\end{figure*}

\newpage

\begin{center}
{\large \Blue{\bf {Galaxy Cluster Shapes}}}

\vspace{0.75cm}

\Red {S. Basilakos$^1,2$, M. Plionis$^1$ \& S. Maddox$^3$}

\vspace{0.5cm}

\Cyan {$^1$
Astronomical Institute, National Observatory of Athens,  \\
I.Metaxa \& Bas.Pavlou, Lofos Koufou, 15236, Athens, Greece
}

\Cyan {$^2$
Section of Astronomy \& Astrophysics, Univ. of Athens,\\
Panepistimioupolis, 15784 Zografos, Athens, Greece
}

\Cyan {$^3$
Institute of Astronomy, Madingley Road, Cambridge, CB3 0EZ, UK}

\end{center}

\vspace{0.5 cm}

\renewcommand{\baselinestretch}{1.3}
\Brown {
We study the performance of two different methods used to define
cluster shapes with final aim to study the projected and intrinsic shapes 
of APM clusters (Dalton {\em et al} 1997).
The $\Red {1^{st}}$ method defines the cluster ellipticity
by fitting ellipses to the individual galaxy distribution as a function
of radius from the cluster centre. The $\Green {2^{nd}}$ method is based on 
smoothing the discrete galaxy distribution and then
fits an ellipse using all cells that fall above an overdensity threshold. This
latter method is free of the known aperture bias which tends to 
artificially sphericalize clusters.
Using Monte-Carlo simulations we have studied the performance of both methods 
in the presence of the expected galaxy background at the different distances 
traced by the APM clusters.}

\renewcommand{\baselinestretch}{1.3}

\begin{center}
{\Red {The Method}} 
\end{center}
In order to estimate the projected cluster shape
we diagonalize the inertia tensor ($\Magenta{\rm det} \Magenta{(I_{ij} - 
\lambda^2 M_{2})=0}$) where $M_{2}$ is the $2 \times 2$ unit 
matrix. The eigenvalues $(a,b)$ with $(a>b)$ define the ellipticity of the
configuration under study: $\varepsilon=1-b/a$ (cf. Plionis, Barrow \& Frenk 
1991).

Initially all galaxy positions are transformed to the coordinate system of 
each cluster by : $x_{i}=(RA_{g}-RA_{cl}) \times \cos(\delta_{cl})$ and 
$y_{i}=\delta_{g}-\delta_{cl}$. Then we use the following procedures:

\noindent
$ \Red {1^{st}}$ \Red{Method}: All galaxies within an initial small radius are 
used to define the initial value of the cluster shape parameters. Then the next
nearest galaxy is added to the initial group and the shape is recalculated.
This method, although straight forward, suffers from the fact that implicitly 
we are assuming a spherical aperture within which the cluster
shape parameters are estimated (cf. Binggeli 1982).

\noindent
$ \Green {2^{nd}}$ \Green{Method}: 
The discrete galaxy distribution is smoothed 
using a Gaussian kernel. Then all cells that have a density above some 
threshold are used to define the moments of inertia with weight
$\Magenta {w_{i}= \rho_{i}-\langle \rho(z) \rangle / \langle \rho(z) \rangle}$
where 
$\Magenta {\langle \rho(z) \rangle =\int_{M_{min}(z)}^{M_{max}} \Phi(M) dM} $
with $\Phi(M)$ the APM luminosity function (with $z$ evolution) from Maddox, 
Efstathiou \& 
Sutherland (1996) and $\Magenta{M_{min}(z)=m_{lim}-5\log(r)-25-3z}$. 
This method is free of the \Blue{aperture bias} and we found that it performs
significantly better than the previous method.

\begin{center}
{\Red { Background Contamination}}
\end{center}
A major problem in reliably determining cluster shapes is the 
significant galaxy background which contaminates the cluster galaxy counts,
especially in deep galaxy catalogues like the APM which has $m_{lim}=20.5$. 
In order to 
assess the effects of such contamination on the determination of cluster shapes
we have performed large sets of Monte-Carlo simulations in which we compared
the determination of cluster shapes, defined to have the King's profile with
core radii $\le 0.1 \; h^{-1}$ Mpc (cf. Girardi {\em et al} 1998), with and 
without the expected background 
contamination. The background at each cluster distance was estimated by:
$$\Magenta {N_{bac}=2\pi(1-cos\theta^{i}) \int_{0}^{z_{max}} \langle \rho(z) 
\rangle z^{2} {\rm d}z}$$ 
where $\theta^{i}$ is the ancular radius of the cluster.
In the figure we show an example of a simulated cluster at 200 $h^{-1}$ Mpc 
with $\varepsilon=0.5$. The performance of the two methods, described above, 
is shown in the same colour coding as in the text. It is evident that the 
$\Green {2^{nd}}$ method performs equally well in the presence or not of the
galaxy background. However, the presence of substructure can affect 
unexpectedly the performance of this method as well.

\begin{figure*}
\mbox{\epsfxsize=15cm \epsffile{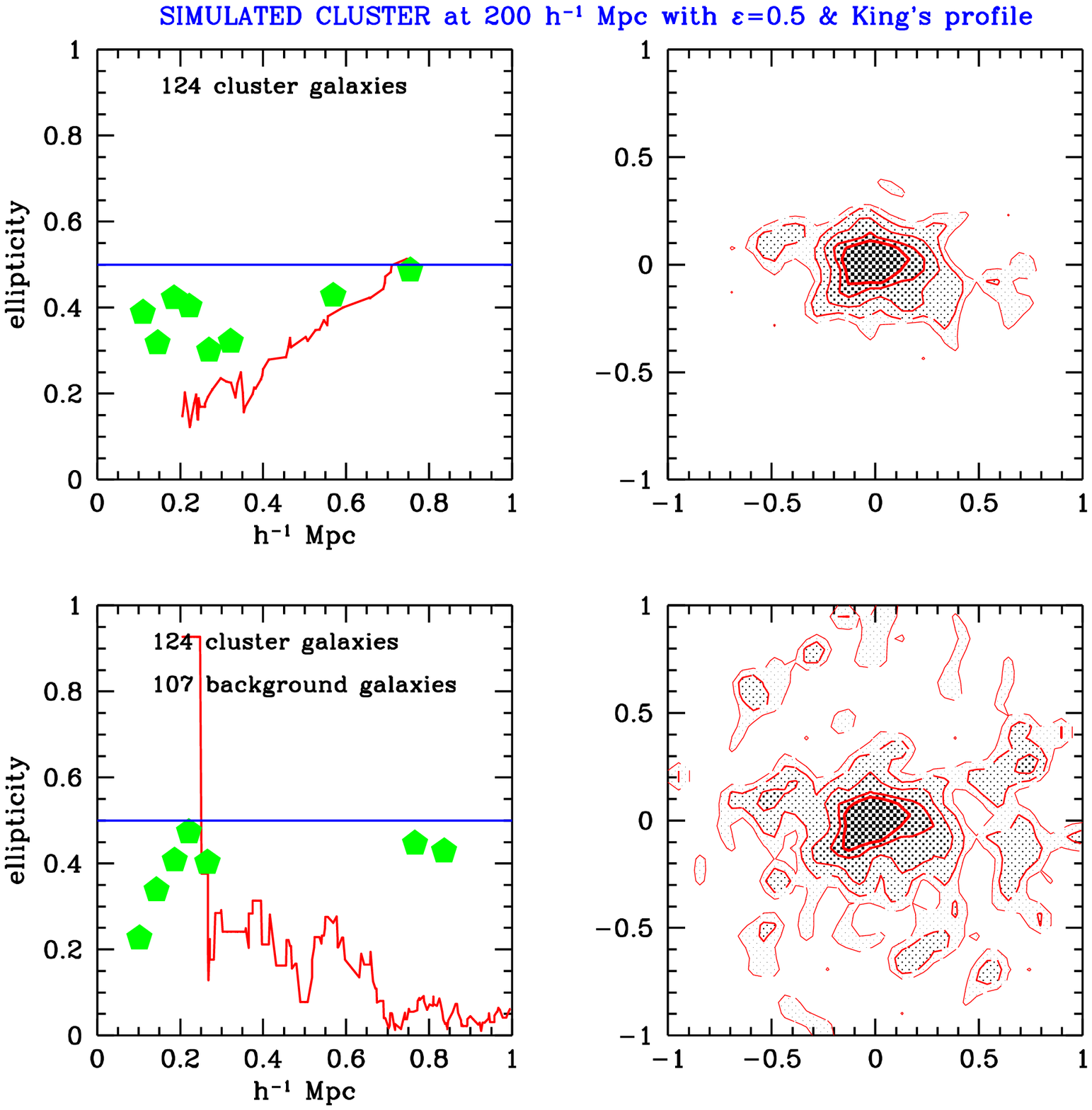}}
\end{figure*}

We are at the process of quantifying biases related to the above methods
and applying our shape determination procedure to the real APM data.

{\small
\renewcommand{\baselinestretch}{1.}

}


{\small
\begin{thebibliography}{}
\bibitem{} Avni, Y., Tananbaum, H., 1986, ApJ, 305, 83
\bibitem{} Fabbiano, G., Kim, D.W., Trinchieri, G., 1992, ApJS, 80, 531
\bibitem{} Ho, L.C. Filippenko, A.V., Sargent, W.L., 1995, ApJS, 98, 477
\bibitem{}    Ho, L.C., Filippenko, A.V., Sargent, W.L., 1997, ApJ, 487, 568 
\bibitem{}  Lahav, O.  et al. 1993, Nature, 364, 693
\bibitem{}  Zezas, A., Georgantopoulos, I. Ward, M.J., 1998,
  Astron. Astroph. Tran., in press 
\end{thebibliography}
}


{\small
\begin{thebibliography}{}
\bibitem{}Iwan D., Marshall F.F., Boldt E.A., Mushotzky R.F., Shafer R.A., 
\& Stottlemyer A., 1982, ApJ, 260, 111
\bibitem{}Kneissl R., Egger R., Hasinger G., Soltan A.M. \& Trumper J.,
1997, AA, 320, 685
\bibitem{}Lahav O., Piran T. \& Treyer, M., 1997, MNRAS, 284, 499
\bibitem{}Shafer R.A., Fabian A.C., in Abell G.O., Chincarini G., eds, IAU
Symp. No 104, Dordrecht, p.33
\bibitem{}Snowden S.L. et al, 1995, ApJ, 454, 643
\end{thebibliography}
}


{\small
\begin{thebibliography}{}
\bibitem{}  N. Bade {\em et al}, 1998, A\&AS, 127, 145
\bibitem{}  B.J. Boyle {\em et al}, 1993, MNRAS, 260, 49
\bibitem{}  F. La Franca, P. Andreani \& S. Christiani, 1998, ApJ, 497, 529 
\bibitem{}  D.Hartmann \& W.B.Burton, 'Atlas of Galactic Neutral Hydrogen', 
1997, Cambridge Univ. Press.
\bibitem{}  T.Shanks \& B.J.Boyle, 1994, MNRAS, 271, 753
\bibitem{}  W. Voges {\em et al}. 1996 IAU Circ., 6420, 2 
\end{thebibliography}
}


\begin{thebibliography}{}
\bibitem{} B. Binggeli, 1982, A\&A, 250, 432
\bibitem{} G.B. Dalton {\em et al}, 1997, MNRAS, 289, 263
\bibitem{} M. Girardi {\em et al.}, 1998, astro-ph/9804187
\bibitem{} S. Maddox, G. Efstathiou, W.J. Sutherland, 1996, MNRAS, 283, 1227
\bibitem{} M. Plionis, J.D. Barrow, C.S. Frenk, 1991, MNRAS, 249, 662
\end{thebibliography}
\end{document}